# Improved Scatter Correction in X-Ray Cone Beam CT with Moving Beam Stop Array Using John's Equation

Hao Yan, Xuanqin Mou, Shaojie Tang, Qiong Xu

*Abstract*—In this paper, an improved scatter correction with moving beam stop array (BSA) for x-ray cone beam (CB) CT is proposed. Firstly, correlation between neighboring CB views is deduced based on John's Equation. Then, correlation-based algorithm is presented to complement the incomplete views by using the redundancy (over-determined information) in CB projections. Finally, combining the algorithm with scatter correction method using moving BSA, where part of primary radiation is blocked and incomplete projections are acquired, an improved correction method is proposed. Effectiveness and robustness is validated by Monte Carlo (MC) simulation with EGSnrc on humanoid phantom.

## I. INTRODUCTION

SCATTER is an open problem in CBCT. Various scatter correction methods are proposed [1-9] and yet continued research is necessary [10]. In the existed methods, scatter estimation with BSA measurements, herein named BSA method [1], is a reliable way. In this method, besides the usual scan, using an extra scan with BSA, scatter is measured view by view. Considering that scatter is slow-variant, it is estimated with 2D spatial interpolation based on the measurements. When the estimated scatter is removed from the projections acquired in the usual scan, primary is got and scatter correction is achieved. Its main limitation is much dose due to the extra scan. Improved BSA method greatly reduces the dose by adopting a sparse-view extra scan with BSA for scatter measurements and estimations. Scatter in other views is estimated by angular cubic interpolation with the estimated scatter in the measured sparse views [2]. For more dose reduction and operation facility, efforts are aiming at integrating the step of scatter measurements into the usual scan, i.e., using appurtenances like moving BSA or collimator leaves to get scatter-removed projections in single scan [3, 4]. Here we focus on the moving BSA method [3], in which scatter is measured in each view with moving BSA and the lost primary blocked by BSA is spatially interpolated. Different with scatter estimation, spatial interpolation has limitation in the estimation of blocked primary, because it performs well only in low frequency but not all the primary is in that way. For this reason, although in [3], the authors design a raster-moving BSA to prevent primary being blocked at fixed position, thereby to reduce the cumulated interpolation error, streak artifacts and noise increase are inevitable in the reconstructed image.

Aiming at overcome above-mentioned limitations, we are searching novel *interpolation* method for restoring more lost information. Based on the seminal work about John's Equation [11], we get a correlation between neighboring CB views. Accordingly, correlation-based algorithm restoring incomplete views is designed. We name it *view-completing* algorithm (VCA). Through combining VCA with moving BSA configuration, we get an improved scatter correction method.

## II. METHODS

### A. John's Equation in CBCT Configuration

Weighted 3D x-ray transform satisfies a cone beam consistency condition, named John's Equation [12]. Denote x-ray spot as $\xi$ and detector cell $\eta$. $g(\xi;\eta)$ is the line integral of object $f$ through $\xi$ and $\eta$,

$$g(\xi;\eta) = \int_{\mathbb{R}} f(\xi + t(\eta - \xi))dt = X(f(\xi;\eta)) \cdot |\xi - \eta|^{-1} \quad (1)$$

$X(f(\xi;\eta))$ is CT data (3D x-ray transform). Denote $g_{xy}$ as the partial differential of $g$ to variables $x$ and $y$. John's Equation is:

$$g_{\eta_i \xi_j}(\xi;\eta) - g_{\eta_j \xi_i}(\xi;\eta) = 0, i, j = 1,2,3 \quad (2)$$

For spiral CBCT shown in Fig. 1, (2) is integrated to (3) [11]:

$$g_{vt} - rg_{uz} = -\frac{2u}{r+d}g_v - \frac{uv}{r+d}g_{vv} - \left(r + d + \frac{u^2}{r+d}\right)g_{uv} \quad (3)$$

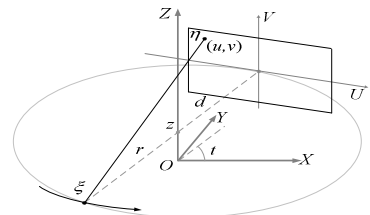

Fig. 1. flat panel CBCT configuration. $u$ and $v$ are coordinates on the 2D detector. $t$ is the rotation angle. $r$ and $d$ represent the distances of source to center and center to detector respectively. $z$ is the longitudinal coordinate

This work is partially supported by National Natural Science Foundation of China (No. 60551003), the fund of the Ministry of Education of China (No.20060698040) and the program of NCET of China (No. NCET-05-0828).

Hao Yan, Xuanqin Mou, Sahojie Tang and Qiong Xu are with the Institute of Image Processing and Pattern Recognition, P.R.China (corresponding author: Xuanqin Mou, phone: 86-29-82663719; e-mail: xqmou@mail.xjtu.edu.cn).

Approximation $g_{uz} \cong g_{uv} \cdot \frac{(r+d)^2 + u^2 + v^2}{r(r+d)}$ is adopted into (3), similarly as that used in [13], we get:

$$g_{vt} \cong -\frac{2u}{r+d} g_v - \frac{uv}{r+d} g_{vv} + \frac{v^2}{r+d} g_{uv} \qquad (4)$$

Denote $g^*(t, k_1, k_2)$ as the Fourier transform of $g(t, u, v)$, wherein superscript * representing Fourier transform. We will simplify $g^*(t, k_1, k_2)$ as $g^*(t)$ and $g(t, u, v)$ as $g(t)$ later. In frequency domain, the counterpart of (4) is:

$$k_2 g^*_t \cong \frac{j}{r+d}\left(-2k_1 g^*_{k_2} + k_2^2 g^*_{k_1 k_2} - k_1 k_2 g^*_{k_2 k_2}\right) \qquad (5)$$

On this basis, we notice that further derivation is:

$$g^*_t \simeq \frac{j}{r+d} \cdot \left(-2\frac{k_1}{k_2} g^*_{k_1} + k_2 g^*_{k_1 k_2} - k_1 g^*_{k_2 k_2}\right), k_2 \neq 0 \qquad (6)$$

Denote the right part of (6) as $G(t, g^*(t))$, $\Delta t$ represents a tiny rotation in $t$, (6) could be simply written as:

$$\frac{g^*(t+\Delta t) - g^*(t)}{\Delta t} \simeq G(t, g^*(t)), k_2 \neq 0.$$

i.e., $g^*(t+\Delta t) \simeq g^*(t) + \Delta t \cdot G(t, g^*(t)), k_2 \neq 0.$ (7)

From (7), we get *a correlation between neighboring CB projections in frequency domain* (*except longitudinal zero frequency*). Using this correlation, it is possible to compute $g^*(t+\Delta t)$ (except $g^*(t+\Delta t, k_1, 0)$) from neighboring view $g^*(t)$. Compared with common spatial interpolation, the correlation in (7) would be a novel promising *interpolation* because it supplies a quasi-exact way utilizing angular-contained information.

*B. View-completing Algorithm (VCA)*

For incomplete spiral CB projections $g(t+\Delta t)$, according to (7), we can develop VCA to restore $g(t+\Delta t)$ using information in the neighboring views $g(t)$ and $g(t+2\Delta t)$.

1) Firstly, $g(t+\Delta t)$ is initially restored by spatial interpolation. The result is denoted as $g^{IR}(t+\Delta t)$, wherein superscript means *initial restoration*.
2) If one of the neighboring views is complete, e.g., $g(t)$ is complete. In frequency domain, $g^*(t)$ is put into (7) and current view $g^{*C}(t+\Delta t)$ is computed, wherein superscript means *computed*. Since (7) is not applicable for $k_2=0$, we just let $g^{*C}(t+\Delta t, k_1, 0)$ equal to $g^{*IR}(t+\Delta t, k_1, 0)$. Back to space domain, we get $g^C(t+\Delta t)$. Corresponding to blocked pixels, value of $g^{IR}(t+\Delta t)$ is replaced by $g^C(t+\Delta t)$ and refined restoration $g^{RR}(t+\Delta t)$ is got, wherein superscript means *refined restoration*.
3) If both neighboring views are incomplete, when compute $g^C(t+\Delta t)$ with $g(t)$, $g(t)$ needs to be spatial interpolated also. The following procedure is the same with 2).

In our opinion, in the situation of 3), although both $g(t)$ and $g(t+\Delta t)$ are incomplete, complementary information exists. In another word, between neighboring incomplete views, one has part of complement information that the other lost, and vice versa. We inferentially realize this since it is well known that there is redundancy in 3D projections. Now we will study whether this is reasonable. Referring to (3), we are aware that John's Equation is local in spiral CBCT. For neighboring CBCT views, according to tiny angular rotation, redundancy is contained in the neighborhood of same position. i.e., *if pixel (u, v) is blocked in current view, to recover it, in the neighboring view, information on pixel (u, v) and its peripheral pixels should be known*. We name it *View-Completing* Condition (VCC). In the situation of 3), VCA is effective when VCC is fulfilled. VCC will be further validated by simulation later.

*C. Improved Scatter Correction with Moving BSA*

In [3], the blocked positions change according to the views, so VCC is fulfilled (on condition that movement of BSA is large enough). In our following improved scatter correction with moving BSA, utilizing VCA is straightforward:

1) One scan with moving BSA is performed. According to each view, scatter is estimated using the measurements with moving BSA and is subtracted from the views. With this step, scatter-free views are got.
2) For the restoration of blocked primary, VCA is iteratively adopted view by view.

Compared with previous version [3], in proposed method, advantages of (7) and spatial interpolation are combined by VCA. Firstly, more lost information especially high frequency information is restored since (7) is effective in most frequency. Secondly, for longitudinal zero frequency which is beyond the ability of (7), VCA keeps spatial interpolation because it could perform well estimation in low frequency.

## III. SIMULATIONS

Both analytical and MC simulations are performed. Firstly, we validate VCC and the effect of VCA with analytical simulation [14] on FORBILD [15], which is a complex head phantom with rich high frequency details. Secondly, to simulate a realistic application and test robustness under noise (quantum noise and inconsistence due to approximate scatter estimation), MC (Egsnrc [16]) simulation on humanoid phantom Zubal [17] is adopted. FDK algorithm is used in reconstruction [18]. In the reconstructed volume, representative slices such as center slice, off-center-slice existing serious blocking are investigated. Simulation configuration is circular CBCT, because it is the special case of spiral and is more commonly used in practice (Fig.2). The application for spiral case is straightforward.

*A. The Validation on VCC*

We validate VCC with CBCT scan on FORBILD (right-top image in Fig.2). Detector is 850 ×200. View Number is 1080. Each view is blocked by raster-moving BSA. BSA has 10×6 blockers and each blocker shades 5×5 pixels. BSA movements are 0,1,2,3,4,6,7 pixels per view. For each view, horizon 1D cubic spline interpolation as in [3] is used and initial restoration is got. On the other hand, refined restoration is got by VCA (the first cycle of computation).The ratio of refined restoration to initial restoration is defined as Relative Error, see (8), wherein, $View_0$ is the ideal non-blocking projections. For each movement, Relative Error is a 1D array with elements of view numbers. Mean Relative Error is defined in (9).

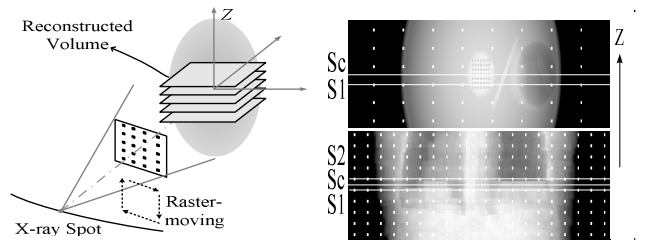

Fig.2. Left: Scanning configuration with moving BSA. Right: One view of Scan. Right top is FORBILD. Right bottom is Zubal For better vision, projections are log operated and elongated in z direction. Investigated slices are marked with white line. Center slice with no blocking is denoted as Sc, and off-center-slices with blocking are denoted as S1, S2.

$$Error_{Relative}(t) = \frac{\sum_u \sum_v |View_{proposed}(u,v,t) - View_0(u,v,t)|}{\sum_u \sum_v |View_{Interpolation}(u,v,t) - View_0(u,v,t)|} \quad (8)$$

$$\overline{Error_{Relative}} = \frac{1}{ViewNumber} \sum_t Error_{Relative}(t) \quad (9)$$

Results are plotted in Fig.3. It demonstrates that when the BSA is static, VCA has little improvement compared with interpolation. When BSA is moving, the positive effect is exhibited. It is more evident with more movements (right image). When movement is large, e.g., 7 pixels per view, the VCC is fully satisfied (recall that blocker shades are 5 pixels width, so 7 pixels are large enough for one blocked area plus a neighborhood). At this time, in each view, VCA is better than interpolation, because in left image, refined restoration is below initial restoration according to each view.

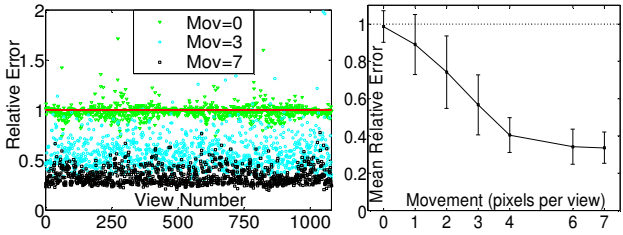

Fig.3. Validation on VCC. Left: Relative Error according to each view with different movements. For better vision, only results of three movements are demonstrated. Right: Mean Relative Error with standard deviation. (In both images, initial restorations are normalized to one (red line in left image and black dotted line in right image)

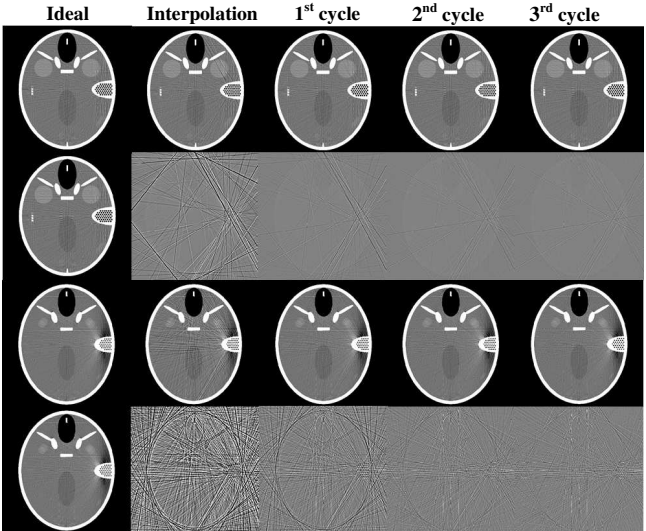

Fig.4. Performance of proposed VCA. Reconstruction of Sc and S1 marked in Fig.2 are displayed in 1st and 3rd row, from left to right respectively is reconstructed image with ideal projection, interpolated projection, results of 1st, 2nd and 3rd cycle of iterations with VCA. Displayed window is [0,100] HU. Accordingly, difference image is shown in 2nd and 4th row, displayed with [-10, 10] HU. In first column of 2nd and 4th row, for better comparison, results of 3rd iteration is duplicated.

### B. Evaluation of VCA with analytical simulation

We iteratively perform *view-completing* to FORBILD with moving BSA. BSA moves 6 pixels per view. The other parameters are same with above section. Results are shown in Fig. 4. It could be observed that streaks due to inaccurate interpolation (2nd column, Fig.4) are greatly reduced with VCA and vision-satisfied restoration could be achieved through three iterations. This evaluation is only with noise-free views. Results for noisy case will be shown in section D where both scatter and quantum-noise are considered.

### C. MC Simulation

To generate scatter data, we simply revise the normal transmission user code in EGSnrc [16]: if photons never been scattered till they will hit the detector, then their weights are set zero. To save time, Richardson-Lucy fitting [19] is adopted. The primary is computed by analytical ray-tracing method [14]. Quantum noise is included in accordance with $10^6$ photons.

To evaluate the accordance between simulation and real equipments, data from micro CT (skyscan 1076) is used. The parameters are the same with [20]. A homogenous water phantom is scanned and result is shown in Fig.5. Profiles of simulated data agree well with real data. The average normalized error is below 5%. Relative large error is observed near the edge for tiny geometry misalignment in the measurement of real data.

### D. Results of Improved Scatter Correction in MC simulation

To eliminate the influence of beam-hardening effect, X-ray source is chosen as monochromatic 60keV (This is common in preliminary MC simulation for scatter correction [3, 5] and will help us study the effect of correction methods alone). Circular chest scan is performed to Zubal phantom [17] (right-bottom image in Fig.2). Distance of source to center and center to detector are 750 *mm* and 375 *mm* respectively. Flat panel (*CsI*) is 800×200, with pixel resolution of 1mm×1mm. Total view number is 540. BSA moves 6 pixels per view. The BSA has 20×10 blockers and the blocker shades 5×5 pixels on the detector. The distance between adjacent blocker shades is 40 and 20 pixels in row and column directions. Considering the penumbra edge effects, scatter is assumed to be measured accurately only in the centered shaded pixel. To combat with scatter (Fig.6), four BSA correction methods are compared (Fig.7). They are BSA method [1], improved BSA methods (the sparse views occupy 5% of the total views) [2], moving BSA methods based on interpolation and proposed moving BSA method with VCA. In comparison, BSA method is referred as an ideal correction. Form Fig. 7, we can see that with our method, evident improvement is achieved compared with method that only uses spatial interpolation, and a quasi-ideal correction is got. In improved BSA method, inaccurate scatter estimation due to sparse-view measuring causes under- or over- correction observed in some positions.

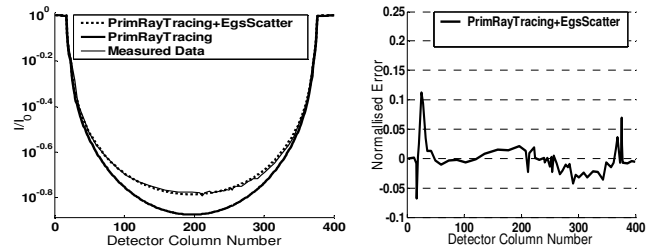

Fig. 5. Validation of MC (Egsnrc) simulation. Left: profiles of measured and simulated projection of water phantom (logarithmic scale). Right: Normalized error of simulation data corresponding to the same profile

## IV. DISCUSSION AND CONCLUSION

In both analytical (Fig.4) and MC (Fig.7) simulations, using proposed methods, significant streaks removing and noise reduction are observed compared with [3]. We reveal that VCA is effective when VCC is fulfilled and validate this with simulation. Considering that VCA works among neighboring

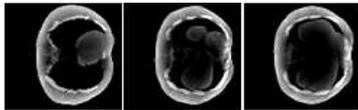

Fig.6. Reconstruction with scatter-polluted views. Window: [-300,200]HU

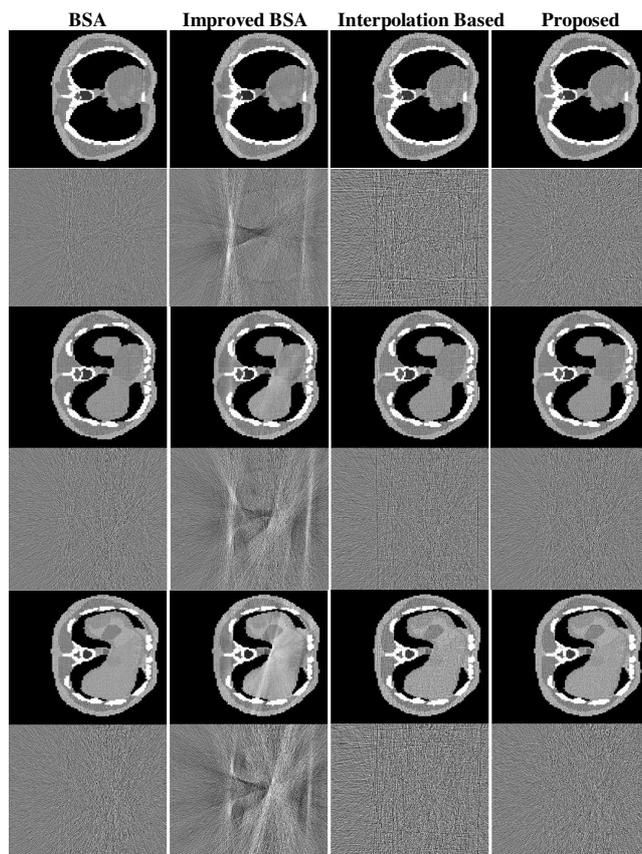

Fig.7. Reconstructions of scatter corrected views. Slices S2, Sc and S1 marked in Fig.2 are displayed. In $1^{st}$, $3^{rd}$ and $5^{th}$ row, from left to right: reconstructed image of correction with BSA method, improved BSA method, moving BSA method with interpolation and proposed method ($3^{rd}$ iteration). Displayed window is [0,200] HU. Accordingly, difference image is show in $2^{nd}$, $4^{th}$ and $6^{th}$ row, displayed with [-50, 50] HU

views, we think it is applicable independent of the moving way of BSA, as long as the movement of BSA is large enough to fulfill VCC. It is useful because in practice, the movements may have tiny geometry misalignment.

To conclude, *View-completing* algorithm is developed and used in scatter correction with moving BSA. The improved method is efficient, robust and with potential use in practice.

### ACKNOWLEDGMENT

The authors thank the assistance of Dr. Iwan Kawrakow with the usage of EGSnrc, also the kindly help of Dr. Wojtek Zbijewski on the measured data, also the help of Dr. I.George Zubal and Dr. L. Zhu with the use of human phantom, as well as the suggestions of the anonymous reviewers.

### REFERENCES

[1] L. A. Love, R. A. Kruger, "Scatter estimation for a digital radiographic system using convolution filtering," Med. Phys. 14,1987, pp.178–185.
[2] R. Ning, X. Tang, and D. Conover, "X-ray scatter correction algorithm for cone beam CT imaging," Med. Phys. 31,2004, pp.195–1202.
[3] L. Zhu, N. Strobel, and R. Fahrig, "X-Ray Scatter Correction for Cone Beam CT Using Moving Blocker Array," Proc. SPIE 5745, pp. 251–258.
[4] J H Siewerdsen et al., "A Simple, Direct Method for X-Ray Scatter Estimation and Correction in Digital Radiography and Cone-Beam CT," Med. Phys. 33,2006, pp.187–97.
[5] L. Zhu, N. Robert Bennett and Rebecca Fahrig, "Scatter Correction Method for X-Ray CT Using Primary Modulation: Theory and Preliminary Results," IEEE Tran. Med. Imag. 25, 2006, pp.1573–1587.
[6] H., Li, R.,Mohan et al., "Scatter kernel estimation with an edge-spread function method for cone-beam computed tomography imaging," Phys. Med. Biol. 53, 2008, pp.6729–6748.
[7] J. Rinkel, L. Gerfault et al., "A New Method for X-Ray Scatter Correction: First Assessment on a Cone-Beam CT Experimental Setup," Phys. Med. Biol. 52, 2007, pp. 4633–4652.
[8] W. Zbijewski, and F. J. Beekman, "Efficient Monte Carlo based scatter artifact reduction in cone-beam micro-CT," IEEE Tran. Med. Imag. 25, 2006, pp.817–827.
[9] Y. Kyriakou, T., Riedel, WA. Kalender, "Combining deterministic and Monte Carlo calculations for fast estimation of scatter intensities in CT," Phys. Med. Biol. 51, 2006, 4567—4586.
[10] Xiaochuan Pan, J. Siewerdsen, P J. La Riviere, WA. Kalender, "University Anniversary Paper: Development of x-ray computed tomography: The role of Medical Physics and AAPM from the 1970s to present," Med. Phys. 35(8), 2008, pp.3729–3739
[11] S. K. Patch, "Computation of Unmeasured Third-Generation VCT Views From Measured Views," IEEE Tran. Med. Imag. 21, 2002, pp.801—813.
[12] F. John, "The Ultrahyperbolic Equation with 4 Independent Variables," Duke Math. J., 1938, pp.300–322.
[13] Michel Defrise, F. Noo and Hiroyuki Kudo, "Improved Two-Dimensional Rebinning of Helical Cone-Beam Computerized Tomography Data Using John's Equation," Inverse Problems 19, 2003, pp.S41--S54
[14] S. Tang, H. Yu, H. Yan, D. Bharkhada, X. Mou, "X-ray projection simulation based on physical imaging model," Journal of X-Ray Science and Technology 14, 2006, pp.177–189
[15] http://www.imp.uni-erlangen.de/forbild/deutsch/results/head/head.html
[16] I.Kawrakow, egspp: the EGSnrc C++ class library, NRCC Report PIRS-899, 2005
[17] Zubal, I.G., Harrell, C.R, Smith, E.O, Rattner, Z., Gindi, G. and Hoffer, P.B, " Computerized three-dimensional segmented human anatomy," Medical Physics, 21(2), 1994, pp. 299–302
[18] L A Feldkamp, L C Davis and J W Kress, "Practical cone-beam algorithm," J. Opt. Soc. Am. 1984, pp.1 612–9.
[19] A.P. Colijn and F.J. Beekman, "Accelerated Simulation of Cone-Beam X-Ray Scatter Projections," IEEE TranMed. Imag.23, 2004, pp.584–590.
[20] A.P. Colijn, A. Zbijewski, Sasov and F.J. Beekman, "Experimental Validation of a Rapid Monte Carlo Based Micro-CT Simulator," Phys. Med. Biol. 49, 2004, pp.4321–4333.